\documentclass[reprint,superscriptaddress,amsmath,amssymb,aps,prl]{revtex4-2}

\usepackage{graphicx}
\usepackage{dcolumn}
\usepackage{bm}
\usepackage{graphicx}
\usepackage{hyperref}
\usepackage{bbm}
\usepackage{xcolor} 
\usepackage{booktabs}
\usepackage[ruled]{algorithm2e}

\hypersetup{
  colorlinks,
  citecolor=magenta,
  linkcolor=magenta,
  urlcolor=magenta}

\def\AS{\textcolor{red}}

\def\scriptI{\mathcal{I}}

\def\be{\begin{equation}}
\def\ee{\end{equation}}

\def\ie{i.e.}

\def\nn{\nonumber}

\def\la{\langle}
\def\ra{\rangle}

\begin{document}
\title{Boosted linear-optical measurements on single-rail qubits with unentangled ancillas}
\author{Aqil Sajjad}
\email{asajjad@umd.edu}
\affiliation{College of Optical Sciences, University of Arizona, Tucson AZ 85721}
\affiliation{Department of Electrical and Computer Engineering, University of Maryland, College Park MD 20742}
\author{Isack Padilla}
\email{iacpad0795@arizona.edu}
\affiliation{College of Optical Sciences, University of Arizona, Tucson AZ 85721}
\author{Saikat Guha}
\email{saikat@umd.edu}
\affiliation{College of Optical Sciences, University of Arizona, Tucson AZ 85721}
\affiliation{Department of Electrical and Computer Engineering, University of Maryland, College Park MD 20742}

\begin{abstract}
Any quantum state of the radiation field, sliced in small non-overlapping space-time bins is a collection of single-rail qubits, each spanning the vacuum and single-photon Fock state of a mode. Quantum logic on these qubits would enable arbitrary measurements on information-bearing light, but is hard due to the lack of strong nonlinearities. With unentangled ancilla single-rail qubits, an $8$-port interferometer and photon detection, we show any single-rail qubit measurement in the $XY$ Bloch plane is realizable with success probability $147/256$, which beats the prior-known $1/2$ limit. 

\end{abstract}

\maketitle
\textit{Introduction.}---There are many encodings of a qubit into bosonic modes, such as the dual-rail, single-rail, GKP, and the cat-basis encoding, each with its own pros and cons in terms of ease of preparation, resilience to photon loss~\cite{Albert2018}, quantum communications capacity~\cite{Noh2019}, interfacing with matter quantum memories~\cite{Dhara2023, Beukers2024}, and the prospects of all-optical universal quantum logic and measurements~\cite{Lund2002}. Photonic quantum information processing is the pursuit of optimal means of encoding, processing and measurement of information-bearing light so as to maximize the efficiency of the information processing task at hand, be it maximizing communications capacity~\cite{Guha_2011}, estimation precision of an entanglement-based photonic sensor~\cite{Tan2008,Xia2020}, or the resolution of an optical imaging system~\cite{Tsang2016b,Grace2021c}. Prescriptions of optimal measurements typically arise from quantum information theory, and appear in the form of abstract operator representations that are hard to translate into a blueprint of a readily realizable system in the modern quantum-optics laboratory. One way to address this is to slice the information-bearing optical pulse or waveform into small bins of non-overlapping space-time modes, since no matter the kind of light involved, e.g., coherent state, multimode squeezed state, thermal state or other, one can arrange the slicing to be fine enough such that each mode is a single-rail photonic qubit, i.e., the qubit encoding in the span of vacuum $|0\ra$, and a single photon Fock state $|1\ra$ of a mode. At this point, any measurement on the original information-bearing light translates into an instruction set on a universal quantum computer acting on single-rail qubits. Single-rail qubits also appear naturally in extremely low photon flux scenarios, e.g., the weak coherent state pulses on the receiver end of a deep-space lasercom~\cite{Guha_2011} or continuous-variable quantum-key distribution~\cite{Leverrier2008}, or the weak multimode thermal light originating from an astronomical source that is well approximated by one photon spread across a large orthogonal temporal mode span such that each mode is a single-rail qubit~\cite{Gottesman2012, Khabiboulline2018,Khabiboulline2019}. Single-rail qubits are also the most natural mediator for entangling emissive atomic qubits~\cite{Barrett2005,Hermans2022}.

Unfortunately, single-rail qubits are very difficult to work with since unitary rotations in the span of $|0\ra$ and $|1\ra$ require highly non-linear operations, resulting in all known quantum gate prescriptions on them being non-deterministic~\cite{Lund2002, Wu2011}. Single photon detection provides a natural $Z$ basis measurement. But measuring a single-rail qubit so as to discriminate between the $X$ basis eigenstates $|\pm\ra \equiv(|0\ra \pm |1\ra)/\sqrt{2}$, or more generally, the 
\begin{equation}
|\pm_\phi\rangle \equiv \frac{1}{\sqrt{2}}(|0\ra \pm e^{i\phi} |1\ra)
\end{equation}
states along an arbitrary azimuthal angle $\phi$ on the $XY$ plane of the qubit Bloch sphere is a major challenge. In the same spirit, we only know how to apply a non-deterministic Hadamard gate~\cite{Lund2002}, which prevents us from carrying out a deterministic $X$ basis measurement by rotating the single-rail qubit to the $Z$ basis.

To bypass these issues, one could consider transferring the quantum state of single-rail qubits on to optically active quantum-logic-capable atomic memories such as color centers~\cite{Hermans2022,Stas2022} or trapped ions~\cite{Inlek2017}, and then apply the quantum logic gates in the latter domain. This philosophy has been explored for instance to derive prescriptions for quantum-optimal laser-light discrimination~\cite{Da_Silva2013}, joint-detection receivers for deep-space laser-light communications~\cite{Rengaswamy2021,Delaney2022,Smith2025}, and long-base imaging telescopes~\cite{ Khabiboulline2018,Khabiboulline2019,Sajjad2023,Padilla2024,Padilla2025}. The well-known teleportation-based state transfer entails first entangling each single-rail photonic qubit with an atomic memory qubit (logical $|0\rangle$ and $|1\rangle$ $\equiv$ $(|\!\!\uparrow\rangle \pm |\!\!\downarrow\rangle)/\sqrt{2}$) by applying a controlled X (CNOT) gate, followed by measuring out the photonic qubit in the $X$ basis, and applying a logical $Z$ gate on the memory qubit if the measurement outcome is $|-\rangle$~\cite{NielsenChuang2010}. The latter step of single-qubit spin control is easy on most atomic qubits. The CNOT gate is realizable by reflecting the single-rail qubit off a cavity-coupled atomic spin qubit in the strong coupling regime that imparts an optical phase when a photon interacts with the atomic qubit in the $|\!\!\downarrow\rangle$ state~\cite{DuanKimble2004}, which has been experimentally realized~\cite{Bhaskar2020}. This makes the single-rail $X$ basis measurement the only bottleneck in the aforesaid state transfer procedure. Being able to optically realize an $X$ measurement on single-rail qubits deterministically, therefore, could open the door to many applications.

A $50\%$ success rate for an $X$ basis measurement on the single-rail qubit $|\psi\ra$ can be achieved by mixing $|\psi\ra$ with a single-rail $|+\rangle$ ancilla in a balanced beam splitter, and measuring the photon numbers at the two outputs. This follows from the intimate relationship between measuring $X$ on a single-rail qubit and that of the $50\%$ success-rate Barrett-Kok partial Bell-state measurement on a pair of single-rail qubits~\cite{Kok2010}, as noted in~\cite{Khabiboulline2019}. If we detect no photons in both output ports, or two in one output and none in the other, then it means picking out the $|0\rangle$ or $|1\rangle$ in our single-rail input $|\psi\rangle$, hence measuring it in the $Z$ basis and failing with the $X$-basis measurement. However, obtaining a single photon in one of the two outputs and none in the other represents a successful $X$ basis measurement. This works because if we only obtain a total of $1$ photon in the outputs, we cannot tell whether it came from $|\psi\rangle$ or the ancilla in the $|+\rangle$ state, hence scrambling the information about where the photon originated from. The beam splitter gives the sum of the single-photon terms in $|\psi\rangle$ and the ancilla in one output, and the difference in the other. With equal amplitudes for the zero and single-photon terms in $|+\rangle$ and $|-\rangle$, a single photon in the output with constructive (destructive) interference signifies $|\psi\rangle$ being measured in the $X$ basis and found to be in $|+\rangle$ ($|-\rangle$). Note that this perfectly constructive and destructive interference only arises with a $|+\rangle$ or $|-\rangle$ single-rail ancilla. 

It turns out however that even with a far-easier-to-produce coherent state ancilla, with amplitude $\alpha=\pm1$, one yields a success rate of about $41.58\%$ for an ideal $X$-basis measurement. This is significant, since without ancilla assistance, $X$-basis measurement on single-rail qubits is not known to work, and the preparation of $|\pm\rangle$ ancilla states is experimentally challenging (see Supplementary Material for further discussion.)
It was suggested in Ref.~\cite{Khabiboulline2019} that one could extend the $|+\rangle$ ancilla boosted scheme to a fully deterministic $X$ measurement by better hiding the which-path information of the photon by sending it through a multiple-port beam splitter circuit such as the Quantum Fourier Transform (QFT) unitary, and employing a large number of $|+\rangle$ ancillas.

\begin{figure}
    \centering   \includegraphics[width=1\linewidth]{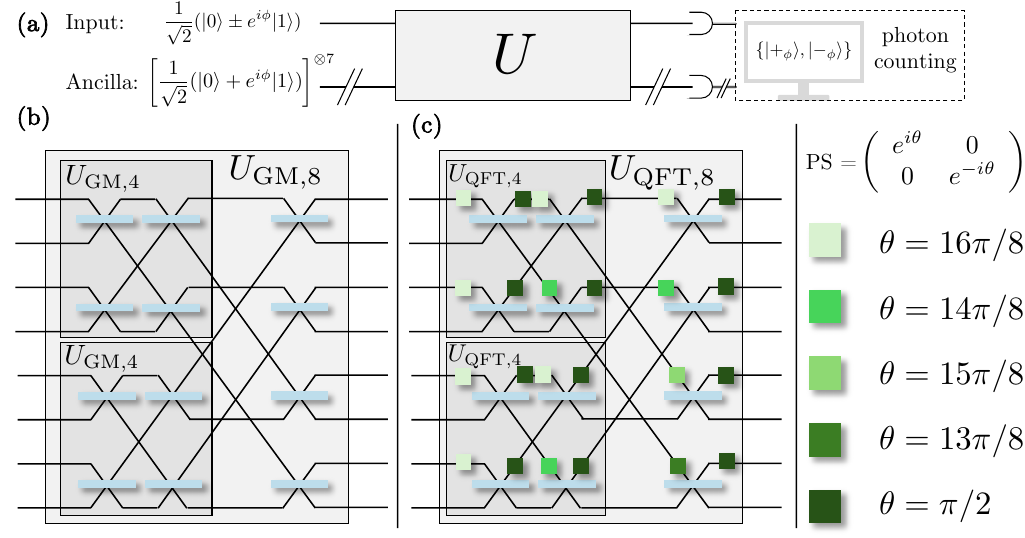}
    \caption{(a) A linear-optical unitary $U$---either an $8$-th order Quantum Fourier Transform (QFT) or an $8$-mode Hadamard `Green Machine' (GM) circuit---receives in the first input mode the single-rail qubit that we wish to measure in the $|\pm_\phi\ra \equiv (|0\ra\pm e^{i\phi} |1\ra)/\sqrt{2}$ basis. Remaining inputs are fed $|+_\phi\ra$ ancillas that help boost the success probability of measuring in the desired basis. (b) The GM circuit $U_{\rm{GM}}$ is realized with a mesh of 50-50 beam splitters and (c) the QFT $U_{\rm{QFT}}$ adds phase shifters, following the construction from Ref.~\cite{Barak2007}.}
    \label{fig:diagram}
\end{figure}

In this paper, we investigate a scheme for boosting the success rate for a single-rail $X$ basis measurement by interfering the single-rail qubit $|\psi\ra$ to be measured, with $n-1$ unentangled single-rail ancillas prepared in the $|+\rangle$ state, in an $n$-mode linear interferometer $U$, followed by photon detection (see Fig.~\ref{fig:diagram}(a)). We then extend this to measuring $|\psi\ra$ in the general $|\pm_\phi\rangle$ basis along any arbitrary direction in the $XY$ plane of the Bloch sphere by employing a collection of $|+_\phi\rangle$ ancilla states. For $U$, we consider both the $n$-port Quantum Fourier Transform (QFT), as well as the $n=2^k \,\,(k \in {\mathbb Z}^+$) mode Hadamard unitary~\cite{Sylvester1867, Paley1933}. 
The power-of-2 Hadamard unitaries admit a particularly simple implementation in terms of $n\log_2(n)/2$ 50-50 beam splitters, termed the `Green Machine' (GM)~\cite{Guha_2011}, without requiring a collection of complex phases as in the QFT unitary of the corresponding size.

We calculate the success probabilities for the above-mentioned approach by considering up to $n=10$ QFTs and $n=8$ Hadamard unitaries, and deduce a general formula for the success rates for general $n$, with the conclusion that the maximum success rate is about $57\%$ and is obtained for the $n=8$ QFT or Green Machine (see Fig.~\ref{fig:diagram}), thereby making this a near-term implementable idea, given the recent experimental realization of the $n=16$ Green Machine for demonstrating super-additive capacity~\cite{Cui2025}. 

Our findings can be summarized as follows:
\begin{enumerate}
\item The power-of-2 Hadamard unitaries yield the same success rate as the QFTs of the same size $n$.
\item For even $n$, the success rate slowly increases from $50\%$ (for $n=2$) up to a maximum value of ${147}/{256} \approx 0.5742$ for $n=8$. On increasing $n$ further, it starts decreasing again, asymptotically falling back to $50\%$ in the large $n$ limit for even $n$.
\item For $n=4$ and $n=6$, we get close to the above-mentioned maximum success rate, i.e., ${9}/{16} =0.5625$ and ${55}/{96} \approx 0.5729$, respectively.
\item For odd $n$, the success rate is ${(n-1)}/{2n}$, which approaches $50\%$ from below for large $n$.
\item The measurement technique is asymmetric. For $n=2$, we have $50\%$ success rate for unambiguously measuring both $|\pm\ra$. But for higher $n$ values, the success rate for measuring $|-\rangle$
slowly rises and approaches $100\%$ for $n\to\infty$. But the probability of successfully measuring a qubit in $|+\ra$ is 0 for odd $n$, and for even $n$, it slowly decreases with $n$ and asymptotically falls to 0 as $n\to\infty$.
\item if we use $|-\ra$ ancillas instead of $|+\ra$, then the success rates for $|+\ra$ and $|-\ra$ are interchanged. Therefore, if we are measuring several copies of a system, we can conduct half of our measurements with $|+\ra$ ancillas and the rest with $|-\ra$ to obtain symmetric results.
\item We prove that the above generalizes straightforwardly to measuring in the $|\pm_\phi\rangle = (|0\rangle \pm e^{i\phi} |1\rangle)/\sqrt{2}$ basis  if we employ $|+_\phi\rangle$ ancillary states instead of $|+\rangle$ ones, and the success rates are independent of the value of $\phi$.
The simpler approach for measuring in the $|\pm_\phi\rangle$ basis, however, would be to apply a $e^{-i\phi}$ optical phase on the single-rail qubit and measure it in the $X$ basis using $|+\rangle$ ancillas.
\end{enumerate}

\textit{Calculating the success and failure rates.}---Consider the set-up for measuring a single-rail qubit in the $|\pm_\phi\rangle$ basis by feeding the optical mode encoding this qubit into the first input of an $n$-component linear interferometer, with the optical modes encoding $|+_\phi\rangle$ ancillary qubits in the remaining input ports. Going forward, we will at times refer to the single-rail qubit and the optical mode encoding it interchangeably, since the meaning will be clear from context. We describe how the $|+_\phi\rangle$ ancillas can be generated in the Supplementary Material. 

The measurement in the $|\pm_\phi\rangle$ basis thus turns into the problem of discriminating between
$|+_\phi\ra^{\otimes n}$ 
and $|-_\phi\ra \otimes |+_\phi\ra^{\otimes (n-1)}$
by feeding these into a linear interferometer and measuring the output photon click patterns.
In terms of the creation operators, these states can be expressed as
\begin{align}
|\pm_{\phi, 1} +_{\phi, 2} +_{\phi, 3} \ldots +_{\phi, n}\ra
&= \frac{1}{2^{n/2}} (1\pm e^{i\phi} a_1 ^\dagger)\nn\\&\quad\times\prod_{j=2}^n(1+e^{i\phi} a_j ^\dagger) |0_1 0_2 \ldots 0_n\ra.
\label{input-state}
\end{align}
Here $|\pm_{\phi, j}\ra$ denotes qubit number $j$ in the state $|\pm_\phi\ra$, and $a_j ^\dagger$ the photon creation operator for the optical mode encoding the $j$-th single-rail qubit.

A linear interferometer transforms the creation operators as $a_j ^\dagger \mapsto \sum_{k=1}^n U_{jk} b_k ^\dagger$,
where $b_k ^\dagger$ are the creation operators in the output, and $U$ the unitary transformation matrix. The output state is then $N_\pm(\phi, b_1 ^\dagger, b_2 ^\dagger,\ldots, b_n ^\dagger) \, |0_1 0_2\ldots 0_n\ra$,
where
\begin{align}
    N_\pm(\phi, b_1 ^\dagger, b_2 ^\dagger , \ldots ,b_n ^\dagger) & \equiv \frac{1}{2^{n/2}} \left(1\pm e^{i\phi} \sum_{k_1 =1}^n U _{1 k_1} b_{k_1} ^\dagger\right) \nn\\
&\quad\times\prod_{j=2}^n \left(1 + e^{i\phi} \sum_{k_i =1}^n U _{j k_i} b_{k_j} ^\dagger\right).
\label{N_pm-def}
\end{align}
The operators $N_\pm(\phi, b_1 ^\dagger, b_2 ^\dagger , \ldots ,b_n ^\dagger)$ are thus polynomials in terms of the output creation operators of the form
{\small
\be
N_\pm(\phi, b_1 ^\dagger, b_2 ^\dagger , \ldots ,b_n ^\dagger)
= \sum_{\substack{i_1,\ldots i_n =0,\\ \sum_{k=1}^n i_k\leq n}}^n
e^{\sum_{k=1}^n i_k\phi} c_{\pm,i_1 i_2\ldots i_n} \, \prod_{j=1}^{n}(b_j ^\dagger)^{i_j}.
\label{polynomial-form}
\ee}

The probability of obtaining the click pattern $|i_1 i_2 i_3\ldots i_n\ra$ on measuring the interferometer outputs is
\be
P_{\pm,i_1 i_2\ldots i_n} = |c_{\pm,i_1 i_2\ldots i_n}|^2 i_1! i_2! \ldots i_n!,
\label{probabilities}
\ee
where the factor of $i_1! i_2! \ldots i_n!$ arises from the fact that $(b_j ^\dagger)^{i_j} |0_j\ra = \sqrt{i_j!} |i_j\ra$.
Thus, these probabilities are clearly independent of the value of $\phi$, and hence we will obtain the same success and failure rates for measuring a single-rail qubit in $|\pm_\phi\rangle$ regardless of the value of $\phi$.
From a practical experimental perspective, it makes more sense to apply a single $e^{-i\phi}$ phase gate on our single-rail qubit to rotate the measurement basis from $|\pm_\phi\ra$ to $\pm\ra$, and measure in the latter by employing a collection of $|+\ra$ ancillary states. In this way, we only need to apply one phase gate on the original single-rail qubit, rather than introducing a $e^{i\phi}$ phase in each of the ancillary qubits.

Note that as $n$ increases, finding the coefficients becomes a very laborious exercise with the number of terms in the polynomials $N_\pm(\phi, b_1 ^\dagger, b_2 ^\dagger,\ldots, b_n ^\dagger)$ scaling exponentially, reminiscent of Boson Sampling~\cite{Aaronson2010}. One relatively simple (though not necessarily fast) way to obtain the probabilities is to use the symbolic manipulation features of a mathematical software such as \emph{Maple}, \emph{Matlab} or \emph{Mathematica} to find the coefficients $c_{\pm,i_1 i_2\ldots i_n}$ as the Taylor series coefficients of the polynomials (\ref{polynomial-form}). 
We can then calculate the probabilities of the various click patterns according to (\ref{probabilities}).
We need to run an iterative routine that goes through all the click patterns individually, and computes their probabilities. If $P_{\pm,i_1 i_2\ldots i_n} = 0$
and $P_{\mp,i_1 i_2\ldots i_n} \neq 0$, then the click pattern $|i_1, i_2,\ldots i_n\rangle$ means our single-rail qubit was unambiguously measured and found to be in $|\mp_\phi\rangle$ in the idealized, noiseless and lossless case.
If both $P_{\pm,i_1 i_2\ldots i_n} \neq 0$, then the click pattern represents failure of measurement in the $|\pm_\phi\rangle$ basis.
We need to add up all the success and failure probabilities associated with the different output click patterns to obtain the overall success rate for measuring our qubit in the $|\pm_\phi\ra$ basis. We describe our algorithm for this purpose in the supplementary material~\cite{supplement}.

As we shall discuss shortly, we find that the success rates are the same for QFTs and the corresponding power of 2 Hadamard unitaries. Therefore, we denote the success and failure rates for measuring a qubit in $|\pm_\phi\rangle$ by
$s_{n\pm}$ and $f_{n\pm}$, respectively, for both QFTs and Hadamard codes, and independent of the angle $\phi$ since we have shown that the probabilities of the output click patterns are independent of $\phi$.
Additionally, to gain further insight into the behavior of the success and failure rates, we can classify the various output states $|i_1 i_2\ldots i_n\ra$ in terms of the total photon count $\scriptI = i_1 +i_2 +\ldots +i_n$ in the output click patterns.
Since a lossless and noiseless linear interferometer conserves the total photon number, we have the same total probability
$P_{n,\scriptI}$
of obtaining one of the click patterns with $\scriptI$ photons as the probability of $\scriptI$ photons in the input state. We can then consider the conditional success and failure rates of our $|\pm_\phi\rangle$ basis measurement given that we obtain a click pattern with $\scriptI$ photons for more insight into the behavior of the results as well as apply various consistency checks. We describe this in more detail in the supplementary material.

\textit{Using $|-\rangle$ ancillas instead of $|+\rangle$ ones.}---Note in Eq.~\eqref{input-state} that if we insert a minus sign in front of each input photon creation operator, i.e. take $a_j ^\dagger \to -a_j ^\dagger$, the input states change from
$|\pm_{\phi, 1} +_{\phi, 2} +_{\phi, 3} \ldots +_{\phi, n}\ra$
to $|\mp_{\phi, 1} -_{\phi, 2} -_{\phi, 3} \ldots -_{\phi, n}\ra$,
corresponding to the use of $|-\rangle$ ancillas with the first qubit flipped. This only changes the output state by resulting in all the output port photon creation operators acquiring minus signs.
The terms in $N_\pm(\phi, b_1 ^\dagger, b_2 ^\dagger , \ldots ,b_n ^\dagger)$ corresponding to even values of the total photon count $\scriptI = i_1+i_2+\ldots i_n$, remain exactly unchanged, whereas those with odd values of $\scriptI$ acquire physically inconsequential minus signs. Thus, the probabilities
$P_{+,i_1 i_2\ldots i_n}$
and $P_{-,i_1 i_2\ldots i_n}$ associated with our single-rail qubit being in $|+\rangle$ and $|-\rangle$, respectively,
are switched if we use $|-\rangle$ ancillas instead of $|+\rangle$ ones. Thus the success rates for measuring $|+\rangle$ and $|-\rangle$ are also switched.

We have also numerically found by trying several examples that randomly switching only some of the ancillas from $|+_\phi\ra$ to $|-_\phi\ra$ while keeping the rest in $|+_\phi\ra$ often results in a reduction in the overall success rate. This is not particularly surprising, since having all the ancillas in either $|+_\phi\ra$ or $|-_\phi\ra$ creates symmetries which are helpful for scrambling our single-rail qubit in a way that improves the success rate. Randomly having some qubits in $|+_\phi\ra$ and others in $|-_\phi\ra$ does not appear to maintain the same scrambling. However, a full study considering all such cases is practically impossible, and we do not explore this topic further.

\textit{QFTs and Hadamard unitaries.}---The unitary transformation matrix for an $n$-component QFT has the entries~\cite{NielsenChuang2010} $(U_{\rm{QFT},n,})_{j,k} \equiv \omega^{(j-1)(k-1)}/\sqrt{n}$, where $\omega = \exp(2i\pi /n)$. The Hadamard matrices, on the other hand, are defined as $n\times n$ matrices comprising entirely of `ones' and `minus ones', and satisfy the conditions $H_n H_n ^\top =nI_{n}$ and $\det(H_n)=\pm \sqrt{n}$, where $I_{n}$ is the $n\times n$ identity matrix. Dividing $H_n$ by $\sqrt{n}$ therefore gives a unitary transformation.
 When $n$ is a power of 2, the Hadamard unitary matrix is defined recursively in terms of the $\frac{n}{2}$ component Hadamard unitary~\cite{Sylvester1867}
\be
U_{\rm{GM}, n} = \frac{1}{\sqrt{2}} \begin{pmatrix}
U_{\rm{GM}, n/2} & U_{\rm{GM}, n/2} \\
U_{\rm{GM}, n/2} & -U_{\rm{GM}, n}
\end{pmatrix},
\ee
for $n=2^j, \,
j\in\{1,2,\ldots\} $, where
\be
U_{\rm{GM}, 2} = \frac{1}{\sqrt{2}} \begin{pmatrix}
1 & 1 \\
1 & -1
\end{pmatrix},
\ee
and is implemented by a balanced beam splitter. In the supplementary material accompanying this work, we describe how larger Hadamard unitaries can be constructed from such beam splitters~\cite{Guha_2011}.
We should mention that Hadamard matrices have also been shown to exist for several numbers other than powers of 2 with the general conjecture that these exist for all multiples of 4~\cite{Paley1933}. However, these do not have a nice recursive form as above, and we only consider the $n=12$ case of this briefly in this work with the finding that it only gives a low success rate of about $36.63\%$ of measuring a single-rail qubit in the $|\pm_\phi\rangle$ basis.

\begin{figure}
    \centering
    \includegraphics[width=1\linewidth]{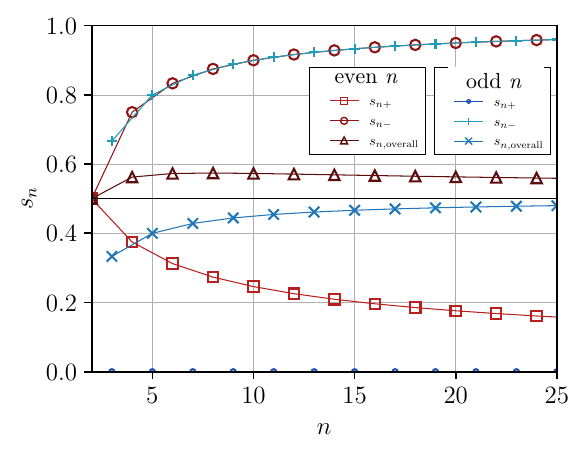}
    \caption{Success probabilities $s_{n}$ of measuring $|\pm_\phi\ra$ as well as the overall success, as a function of the number of inputs  $n$ of the Green Machine (only applies for $n=2^j$, $j\in\{1,2,3,\dots\}$) and QFT (all values of $n$ apply). The maximum overall success is found to be $s_{8,\text{overall}}=147/256$ or $\sim 57.42\%$.}
    \label{fig:plot}
\end{figure}

\textit{Results.}---We have calculated the success and failure rates for QFTs up to $n=10$ and the power-of-2 Hadamard unitaries up to $n=8$. These results have the following patterns (shown in Fig.~\ref{fig:plot}):
\begin{enumerate}
\item For even $n$, the success rate for measuring $|+_\phi\ra$ is
\be
s_{n+} = \frac{n!}{2^n [(n/2)!]^2},
\ee
 which decreases with $n$ and goes to $0$ as $n\to\infty$. For odd $n$, the success rate for measuring $|+_\phi\ra$ is exactly zero for all $n$.
\item The success rate for measuring $|-_\phi\ra$ for all (even or odd) $n$ is
\be
s_{n-} = \frac{n-1}{n},
\ee
which approaches unity as $n\to\infty$.
\item For even $n$, the overall success rate (which is the average of success rates for $|+_\phi\ra$ and $|-_\phi\ra$ if we assume equal prior probabilities for both,) is then given by
\be
s_{n, \rm overall} = \frac{1}{2}\left(\frac{n!}{2^n [(n/2)!]^2} +\frac{n-1}{n}\right) \,\, \text{for even} \, n
\ee
This is equal to $0.5$ for $n=2$, ${9}/{16} = 0.5625$ for $n=4$, ${55}/{96} \approx 0.5729$ for $n=6$,
and reaches a maximum of ${147}/{256} \approx 0.5742$ for $n=8$. After that, it slowly decreasing and asymptotically approaches $50\%$ from above as $n\to\infty$ instead of increasing and approaching unity as speculated by~\cite{Khabiboulline2019}. For odd $n$, the overall success rate (again, assuming equal priors,) is
\be
s_{n, \rm overall} = \frac{n-1}{2n} \,\, \text{for odd} \, n
\ee 
This asymptotically approaches $50\%$ from below as $n\to\infty$.
\end{enumerate}
Some further observations based on the break-down in terms of the total number of photons in the various output click patterns are given in the supplementary material~\cite{supplement}.

The fact that the success and failure rates are asymmetric between $|+_\phi\ra$ and $|-_\phi\ra$ is not entirely surprising given that the ancillas are all $|+_\phi\rangle$. Recall our earlier discussion that if we use $|-_\phi\rangle$ for all the ancillas, then the success rates for $|+_\phi\ra$ and $|-_\phi\ra$ are swapped. Therefore, in a setting where we measure multiple copies of a single-rail photon in the $|\pm_\phi\ra$ basis, it may be worth carrying out half of the measurements with $|+_\phi\ra$ ancillas, and the other half with $|-_\phi\ra$ ancillas.
However, as mentioned earlier, we have also briefly considered using $|+_\phi\rangle$ for some of the ancillas, and and $|-_\phi\rangle$ for the rest, and have found that it does not lead to any improvement and in some cases even results in reducing the success rates, though we do not have a rigorous proof and cannot rule out possible exceptions with $100\%$ confidence.
 
\textit{Conclusion.}---We have investigated the problem of measuring a single-rail qubit in the $|\pm_\phi\rangle$ basis along an arbitrary axis of the $XY$ plane of the qubit Bloch sphere by employing $n-1$ unentangled single-rail ancillas in the $|+_\phi\rangle$ state, along with an $n$-component QFT or the power-of-2 Green Machine realizing the Hadamard unitary. Our proposed prescription is to feed the optical mode encoding the single-rail qubit into the first input port of a QFT (or a power-of-2 Green Machine), with single-rail $|+_\phi\rangle$ ancilla states fed into the other inputs, and detect photons in all outputs using photon number resolving detectors. By calculating the success rates for up to $n=10$, we observe the patterns and deduce a conjecture for the general formula for the success rate for arbitrary $n$. We obtain the same success rate for an $n$-component Green Machine and the QFT when $n$ is a power of 2, and prove that it is independent of the value of the azimuthal angle $\phi$. We find that this strategy boosts the measurement success rate up to a maximum of $147/256\approx 57.42\%$ for $n=8$, but does not give a fully deterministic measurement that Ref.~\cite{Khabiboulline2019} conjectured. Moreover, the simpler $n=4$ Green Machine or QFT comes fairly close to this maximum value with a success rate of $9/16 = 56.25\%$, and is readily realizable~\cite{Cui2025}. 

The measurement strategy is, however, very asymmetric. Except the $50\%$ success probability for $n=2$, the probabilities for successfully measuring a qubit that is in $|+_\phi\ra$ and $|-_\phi\ra$ are uneven. The success rates for the latter are ${n}/{(n-1)}$ for all (even and odd) $n$. But those for the former slowly fall to zero for large $n$ if $n$ is even, and are exactly 0 for all odd values of $n$.
Thus in the large $n$ limit, the probabilities for successfully measuring $|+_\phi\ra$ and $|-_\phi\ra$ approach zero and $100\%$, respectively. Therefore, if we have an application where we care more about measuring $|-_\phi\ra$, then this can still be a fairly useful strategy. On the other hand, for more general applications where we need to measure both  $|+_\phi\ra$ and $|-_\phi\ra$, this method will be inadequate due to its falling success rate for $|+_\phi\ra$.
That said, we have also shown that if we use $|-\ra$ ancillas instead of $|+\ra$ ones, then the success rates of measuring $|+_\phi\ra$ and $|-_\phi\ra$ are simply swapped.
Therefore, we can still benefit from the gain above $50\%$ for $n>2$ in settings where we are measuring multiple copies of the same quantum system by employing $|+_\phi\ra$ ancillas for half of the trials and $|-_\phi\ra$ for the rest.

The fact that we do not obtain a $100\%$ success rate, however, leaves open how to carry out a deterministic single-rail measurement along some direction on the $XY$ plane of the Bloch sphere such as the $X$ or $Y$ bases. Two solutions that attain $100\%$ success rates in the idealized noiseless case will be presented in a separate work~\cite{Sajjad2025b}, but these involve entangled ancillas. Therefore, in the presence of resource constraints, the simplicity of a 50-50 beam splitter or an $n=4$ Green Machine with $|+_\phi\ra$ or $|-_\phi\ra$ ancillas remains an attractive practical option.

It would be remiss not to mention a long line of literature on Bell-basis measurements on dual-rail photonic qubits, i.e., a qubit encoded by the presence of a single photon in one of two orthogonal modes: a no-go theorem that linear-optics alone cannot surpass a $50\%$ success rate~\cite{Calsamiglia2001}, boosting the success rate to up to $25/32$ using unentangled single-photon ancillas~\cite{Ewert2014} and to $59.6\%$ using in-line squeezing~\cite{Zaidi2013,Kilmer2019}, and recent extensions to logical Bell measurements~\cite{Hilaire2023} and experiments~\cite{Hauser2025}. Beating this $50\%$ limit has had profound implications to the resource efficiency of photonic cluster state generation via percolation theory~\cite{Gimeno-Segovia2015,Pant2019}. Given the close relationship between boosting partial Bell-state analysers and single-qubit measurements as discussed above, and the close synergies of the evolution of multi-photon entanglement in linear interferometers, one may be able to borrow ideas from this literature into the single-rail qubit domain.

\textit{The authors are truly grateful to Johannes Borregaard for seeding the original idea during a discussion with SG at The Hague, and for suggesting the use of coherent-state ancillas. The authors thank Joseph Gabriel Richardson for many helpful conversations throughout the course of this work. AS and SG acknowledge funding support from the DARPA Phenom project awarded under contract\# HR00112490451, and the ARO Quantum Network Science MURI awarded under grant\# W911NF2110325. IP and SG acknowledge funding support from AFOSR grant\# FA9550-22-1-0180.}


\bibliographystyle{apsrev4-2}
\bibliography{bibliography}


\clearpage

\onecolumngrid

\section{Supplementary material}

\subsection{Motivation: the state transfer protocol and $X$ basis measurement}
\label{motivation-section}

Consider a single-rail qubit in some arbitrary state $\upsilon |0_p\rangle +\xi |1_p\rangle$, which we wish to transfer to an atomic memory. Following the circuit teleportation protocol~\cite{NielsenChuang2010}, we can bring in an atomic qubit initialized in
$|\overline{0}_a\rangle$. A CNOT gate from the single-rail qubit to the atomic one entangles the two qubits, giving the joint state
$\upsilon |0_p \overline{0}_a\rangle +\xi |1_p \overline{1}_a\rangle$.
It is straightforward to see that an $X$ basis measurement of the photonic qubit leaves the atomic qubit in
$\upsilon |\overline{0}_a\rangle \pm\xi |\overline{1}_a\rangle$, corresponding to the outcome $|\pm_p\rangle$.
Thus, the state transfer is complete if the measurement outcome is $|+_p\rangle$, whereas we need to apply a corrective $Z$ gate if we obtain $|-_p\rangle$.
In theory, this protocol can be implemented using the Duan-Kimble method~\cite{DuanKimble2004}, where the atomic qubit is defined in terms of the spin states of an atom trapped inside a cavity, and the optical mode encoding the single-rail qubit is allowed to interact with this atom. If there is a photon, it reflects off of the $|\downarrow\rangle$ spin state, imparting a $\pi$ phase to it, while leaving the $|\uparrow\rangle$ state invariant. Defining the states $(|\uparrow\rangle + |\downarrow\rangle)/\sqrt{2}$
and $(|\uparrow\rangle - |\downarrow\rangle)/\sqrt{2}$
to be the logical $|0_a\rangle$ and $|1_a\rangle$ states of the atomic qubit, respectively, this arrangement gives us a CNOT gate from the single-rail qubit to the atomic one. An experimental demonstration of such an interaction of an optical mode with an atom in a cavity has been presented in~\cite{Bhaskar2020} in the context of dual-rail qubits, but realized with weak-coherent state approximate time-bin dual-rail qubits.

\subsection{Generating single-rail $|+\ra$ or $|+_\phi\ra$ ancilla qubits}

We can generate a single-rail $|+\ra$ or more generally, a $|+_\phi\ra$ state with color centers such as Nitrogen Vacancy (NV) or Silicon Vacancy (SiV) centers, by using a variant of the protocol described in~\cite{Vasconcelos2020} to produce dual-rail $|+\ra$ states.
A color center effectively behaves as a two-level atom in which the ground state splits into the spin up $|\uparrow_s\ra$ and spin down $|\downarrow_s\ra$ states in the presence of a static magnetic field. The excited state also splits into its spin-up and spin-down versions $|\uparrow^\prime\ra$ and $|\downarrow^\prime\ra$, and the energy gap between them is different from that between the two spin states of the ground state. The system is initialized in a tensor product of the `spin down' atomic state and the vacuum photonic state $|\downarrow_s\ra |0_p\ra$, followed by a $\pi/2$ microwave pulse on the former to have the superposition 
\begin{align}
|\gamma_{sp}\ra =\frac{1}{\sqrt{2}}(|\uparrow_s\ra |0_p\ra +|\downarrow_s\ra |0_p\ra).
\end{align}
We can then apply a pulse resonant with the transition $|\downarrow\ra\to|\downarrow^\prime\ra$. As the atom falls back to the ground state $|\downarrow_s\ra$, a photon is emitted, resulting in the spin-photon entangled state
\begin{align}
|\gamma'_{sp}\ra &=\frac{1}{\sqrt{2}}(|\uparrow_s\ra |0_p\ra 
+|\downarrow_s\ra |1_p\ra) \nn \\
&=\frac{1}{\sqrt{2}}(|+_s\ra |+_p\ra 
+|-_s\ra |-_p\ra) \nn \\
\end{align}
where $|\pm_s\ra\equiv (|\uparrow_s\ra \pm |\downarrow_s\ra)/\sqrt{2}$ are the $X$ basis states of the spin qubit.
Measuring out the atomic qubit in the $X$ basis thus gives us a single-rail $|+\ra$ state if the outcome is $|+_s\ra$. If the outcome is $|-_s\ra$, then we obtain a single-rail $|-\ra$ state, which we can flip into a $|+\ra$ by applying a $Z$ operation.

In the same spirit, we can apply a $\pm e^{i\phi}$ phase delay on a $|\pm\ra$ single-rail qubit to turn it into a $|+_\phi\ra$ state.
Alternatively, we can apply a phase shift gate with the phase $e^{i\phi}$
on the atomic qubit in $|\gamma'_{sp}\ra$, so that $|\gamma'_{sp}\ra$
turns into
\begin{align}
|\gamma'_{sp, \phi}\ra =\frac{1}{\sqrt{2}}(|\uparrow_s\ra |0_p\ra 
+e^{i\phi} |\downarrow_s\ra |1_p\ra).
\end{align}
Rewriting this state in the spin $X$ basis,
we obtain
\begin{align}
|\gamma'_{sp, \phi}\ra=\frac{1}{\sqrt{2}}(|+_s\ra |+_{p, \phi}\ra 
+|-_s\ra |-_{p, \phi}\ra).
\end{align}
where $|\pm_{p, \phi}\ra$
deis the photonic $|\pm_\phi\ra$ state.
Measuring out the spin system in the $X$ basis leaves the photonic qubit in the $|+_\phi\ra$ state if the result is $|+_s\ra$ and $|-_\phi\ra$ if the result is $|-_s\ra$.

\subsection{Implementing power of 2 Hadamard unitaries}
\label{Green-machine-construction}

\begin{figure}
    \centering   \includegraphics[width=0.7\linewidth]{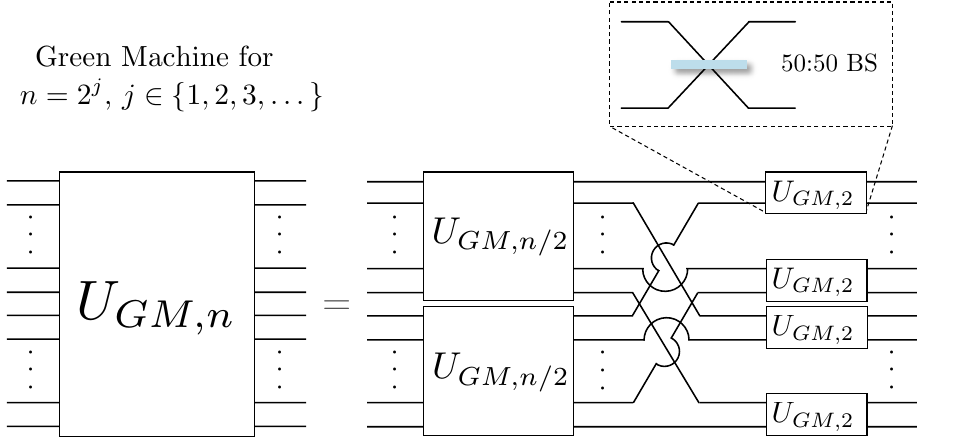}
    \caption{The Green Machine transformation for size $n$ being powers-of-two can be recursively constructed from Green Machines of size $n=2$ (the balanced beam splitter) and size $n/2$.}
    \label{fig:diagram2}
\end{figure}

Note that the 2-component Hadamard unitary $U_{\rm{GM}, 2}$ is implemented by a beam splitter with the transformation
\begin{align}
a_1 ^\dagger &\mapsto\frac{1}{\sqrt{2}} (b_1 ^\dagger +b_2 ^\dagger), \quad
a_2 ^\dagger \mapsto \frac{1}{\sqrt{2}}(b_1 ^\dagger -b_2 ^\dagger)
\label{beam-splitter-convention}
\end{align}
between the input creation operators $a_1 ^\dagger$ $a_2 ^\dagger$ and the output creation operators $b_1 ^\dagger$ and $b_2 ^\dagger$. An $n$-component Green Machine can be built recursively with such 50-50 beam splitters (see Fig.~\ref{fig:diagram2}) as follows:
\begin{enumerate}
\item Apply an ${n}/{2}$-component Green Machine on the first ${n}/{2}$ ports, and another ${n}/{2}$ component Green Machine on the last ${n}/{2}$ ports. 
\item Apply a 50-50 beam splitter of the form (\ref{beam-splitter-convention}) to mix port 1 with port ${n}/{2}+1$. Another such beam splitter to mix port 2 with ${n}/{2}+2$, and so on. Thus for each $j$ from 1 to ${n}/{2}$, we mix port $j$ with ${n}/{2}+j$ with a 50-50 beam splitter.
\end{enumerate}
It is worth specifying that this recursive construction only works when the beam splitters are of the form (\ref{beam-splitter-convention}). It will not work if we have the imaginary number $i$ in the $2\times 2$ transformation. 

\subsection{Breaking down the success and failure rates in terms of the photon counts of the click patterns}

Before we describe our algorithm for calculating the success and failure rates, we describe how we will break down the analysis in terms of the total photon count $\scriptI = i_1 +i_2 +\ldots +i_n$ for the output photon click patterns $|i_1 i_2 \ldots i_n\rangle$. Recall that we are denoting the success and failure rates for measuring a qubit in $|\pm_\phi\rangle$ by
$s_{n\pm}$ and $f_{n\pm}$, respectively.
Naturally, the sum of the probabilities being unity is a nice consistency check on our results:
\begin{align}
\sum_{i_1, i_2\ldots i_n=1}^n P_{\pm,i_1 i_2\ldots i_n} &= 1 \quad\text{and}\quad
s_{n\pm} +f_{n\pm} = 1.
\end{align}
Additionally, we can gain further insight into the behavior of the success and failure rates by classifying the various output states $|i_1 i_2\ldots i_n\ra$ in terms of the total photon count $\scriptI = i_1 +i_2 +\ldots +i_n$ in the click patterns. Since the total number of photons is conserved in a lossless and noiseless linear interferometer, the total probability of output states with $\scriptI$ total photons will be the same as the probability of $\scriptI$ photons in the input states
$|\pm_\phi\ra \otimes |+_\phi\ra^{\otimes (n-1)}$.
Specifically, if we write out each qubit from these input states in the $Z$ basis, we obtain a total of $2^n$ terms, of which $\binom{n}{\scriptI} = n!/[\scriptI! (n-\scriptI)!]$ will have a total of $\scriptI$ photons going in.
These will have the total probability ${\binom{n}{\scriptI}}/{2^n}$,
which will be equal to the total probability of output states with a total of $\scriptI$ photons:
\begin{align}
    P_{n,\scriptI} &\equiv \frac{1}{2^n}\binom{n}{\scriptI} \nn \\
    &= \frac{n!}{\scriptI! (n-\scriptI)! 2^n} \nn \\
    &= \sum_{\substack{i_1, i_2,\ldots ,i_n =1,\\ i_1+i_2+\ldots +i_n =\scriptI}}^n P_{\pm,i_1 i_2\ldots i_n}
\label{P_i-formula}
\end{align}

Now, let $s_{n, \scriptI\pm}$ be the probability of success of measuring and finding our input qubit 1 to be in $|\pm_\phi\ra$ with output click patterns that have a total of $\scriptI$ photons in all output ports, and $f_{n, \scriptI\pm}$ be the corresponding failure probability. Then
\be
P_{n,\scriptI} = s_{n, \scriptI\pm} + f_{n, \scriptI\pm},
\label{photon-number-probability-breakup}
\ee
and
\begin{align}
s_{n\pm} &= \sum_{\scriptI=0}^n s_{n, \scriptI\pm}, \quad
f_{n\pm} = \sum_{\scriptI=0}^n f_{n, \scriptI\pm}.
\label{success-and-failure-rate-photon-number-breakup}
\end{align}
This way, (\ref{photon-number-probability-breakup}) and (\ref{success-and-failure-rate-photon-number-breakup}) provide additional consistency checks on our calculation of the probabilities of the various output click patterns as well as the success and failure rates for measuring our single rail qubit in $|+\ra$ and $|-\ra$.
But more importantly, we will be analyzing the break-up of the success and failure rates in terms of the total photon counts in the output click patterns to deduce the general formulae for the success and failure rates.

\subsection{Our algorithm for calculating the success and failure rates}
\label{algorithm-appendix}

First, recall that the probabilities for the output click patterns can be obtained by writing down the polynomials (\ref{polynomial-form}) in terms of the output port creation operators and then obtaining the coefficients $c_{\pm,i_1 i_2\ldots i_n}$ from a Taylor series command in a computer algebra software like \emph{Maple}, \emph{Mathematica} or \emph{Matlab}. The probabilities for the individual click patterns are then given in terms of the coefficients by (\ref{probabilities}). Since these probabilities are independent of the value of $\phi$, we will work with the $\phi=0$ case to simplify the calculation and avoid having to deal with complex phases.

To determine the success rates from the probabilities, we can use the following procedure (summarized in algorithm~\ref{algorithm:calc_probs}):
\begin{enumerate}
\item To add up the total success and failure probabilities, we introduce variables $s_{n,\scriptI\pm}$ and $f_{n,\scriptI\pm}$ and initialize them with values of 0 before starting any of the loops.
\item We already know that the 0 and $n$-photon click patterns give zero success rate. Hence,
\begin{align}
f_{n,0\pm} &=P_{n,0} \nn \\
f_{n,n\pm} &=P_{n,n},
\end{align}
which we just assign in a single line instead of going through these click patterns.
\item Define an array of variables $[i_1, i_2,\ldots ,i_n]$.
\item Run a set of nested loops over each of these variables $i_1, i_2,\ldots i_n$ etc. where each of them goes from $0$ to $n-1$.
This will allow us to go through each of the output click patterns $|i_1 i_2\ldots i_n\ra$.
\item To avoid going over configurations $[i_1, i_2,\ldots ,i_n]$ for which $\scriptI = \sum_{j=1}^n i_j \geq n$,
we introduce the condition in the $k$-th loop for $2\leq k \leq n$ that checks whether
\be
\sum_{j=1}^k i_j =n
\ee
If this condition is satisfied, then we have a break command, so that we break out of the $i_k$ loop and jump to the next iteration of $i_{k-1}$.
To express this more explicitly, we have $i_1$ going from $0$ to $n-1$. Then inside this, a second loop runs over $i_2$ from $0$ to $n-1$. Inside that, we check whether $i_1+i_2 =n$. If this is true, then we execute the break command for the loop over $i_2$.
If this condition is not satisfied, then we go into the next loop over $i_3$ going from $0$ to $n-1$. Then we check whether $i_1+i_2+i_3=n$. If true, then break, otherwise we go into the $i_4$ loop, and so on.
The purpose of this set of break conditions is that we know that there is zero success rate for any click patterns with $n$ total photons, and there are no click patterns with more than $n$ photons. Therefore, we only need to calculate the probabilities for the cases where
$\scriptI < n$.
\item In the innermost loop, which involves $i_n$ running from $0$ to $n-1$, we assign
$\scriptI = i_1+i_2+\ldots +i_n$.
Then, as described in the previous step, we check whether $\scriptI =n$ and apply a break condition if this is true as described above.
But in the ``else" case, calculate the probabilities $P_{\pm,i_1 i_2\ldots i_n}$ for the click pattern $|i_1 i_2 \ldots i_n\ra$
based on equations (\ref{polynomial-form}) (by employing the Taylor series coefficient command in a mathematical software) and (\ref{probabilities}). We perform these calculations by taking $\phi=0$ for convenience, as we have shown that the results are the same for all values of $\phi$.
\item we then apply the condition that if $P_{+,i_1 i_2\ldots i_n} P_{-,i_1 i_2\ldots i_n} \neq 0$ (i.e. if both are simultaneously non-zero),
then
\begin{align}
f_{n,\scriptI+} &= f_{n,\scriptI+} +P_{+,i_1 i_2\ldots i_n}, \nn \\
f_{n,\scriptI-} &= f_{n, \scriptI-} +P_{-,i_1 i_2\ldots i_n}.
\end{align}
That is, failure for measuring both $|+\ra$ and $|-\ra$.
\item In the ``else" case, \ie~$P_{+,i_1 i_2\ldots i_n} P_{-,i_1 i_2\ldots i_n} =0$, we execute the commands
\begin{align}
s_{n,\scriptI+} &= s_{n,\scriptI+} +P_{+,i_1 i_2\ldots i_n}, \nn \\
s_{n,\scriptI-} &= s_{n, \scriptI-} +P_{-,i_1 i_2\ldots i_n}.
\end{align}
This way, whichever one of $P_{+,i_1 i_2\ldots i_n}$
and $P_{-,i_1 i_2\ldots i_n}$ is non-zero gets added to the corresponding success rate, while the other one which is zero will naturally add nothing. If both are zero, then again, these commands will add nothing to the success rate variables $s_{n,\scriptI\pm}$.
\item We now have all the success and failure probabilities $s_{n,\scriptI\pm}$ and $f_{n,\scriptI\pm}$ for $\scriptI$ total photon clicks for all $\scriptI \in \{0,1,\ldots ,n\}$.
We just need to add them to obtain the total success and failure rates.
\begin{align}
s_{n\pm} &= \sum_{\scriptI=0}^n s_{n,\scriptI\pm}, \nn \\
f_{n\pm} &= \sum_{\scriptI=0}^n f_{n,\scriptI\pm}.
\end{align} 
 \item To make sure that our code is working correctly, we can apply the consistency checks
\begin{align}
s_{n,\scriptI\pm} + f_{n,\scriptI\pm} &= P_{n,\scriptI}, \nn \\
s_{n\pm} +f_{n\pm} &= 1.
\end{align}
\end{enumerate}

\begin{algorithm}[H]
    \caption{Calculation of Success and Failure Probabilities}
    \label{algorithm:calc_probs}
    
    Initialize $s_{n,\mathcal{I}\pm} \gets 0$ and $f_{n,\mathcal{I}\pm} \gets 0$,\quad $\mathcal{I} \in \{0,1,\ldots,n\}$\;
    $f_{n,0\pm} \gets P_{n,0}$\;
    $f_{n,n\pm} \gets P_{n,n}$\;
    Define index array $[i_1, i_2, \ldots, i_n]$\;
    
    \For{$i_1 \gets 0$ \KwTo $n-1$}{
      \For{$i_2 \gets 0$ \KwTo $n-1$}{
        \lIf{$i_1 + i_2 = n$}{\textbf{break}}
        \For{$i_3 \gets 0$ \KwTo $n-1$}{
          \lIf{$i_1 + i_2 + i_3 = n$}{\textbf{break}}
          \DontPrintSemicolon
            $\vdots$\;
          \PrintSemicolon
          \For{$i_n \gets 0$ \KwTo $n-1$}{
            $\mathcal{I} \gets i_1 + i_2 + \cdots + i_n$\;
            \lIf{$\mathcal{I} = n$}{\textbf{break}}
            Calculate $P_{\pm,\, i_1 i_2 \ldots i_n}$ using Eqs.~\eqref{polynomial-form} and \eqref{probabilities}\;
            \eIf{$P_{+,\,i_1 i_2 \ldots i_n} \cdot P_{-,\,i_1 i_2 \ldots i_n} \neq 0$}{
              $f_{n,\mathcal{I}+} \gets f_{n,\mathcal{I}+} + P_{+,\,i_1 i_2 \ldots i_n}$\;
              $f_{n,\mathcal{I}-} \gets f_{n,\mathcal{I}-} + P_{-,\,i_1 i_2 \ldots i_n}$\;
            }{
              $s_{n,\mathcal{I}+} \gets s_{n,\mathcal{I}+} + P_{+,\,i_1 i_2 \ldots i_n}$\;
              $s_{n,\mathcal{I}-} \gets s_{n,\mathcal{I}-} + P_{-,\,i_1 i_2 \ldots i_n}$\;
            }
          }
        }
      }
    }
    $s_{n\pm} \gets \displaystyle\sum_{\mathcal{I}=0}^{n} s_{n,\mathcal{I}\pm}$\;
    $f_{n\pm} \gets \displaystyle\sum_{\mathcal{I}=0}^{n} f_{n,\mathcal{I}\pm}$\;
    \textbf{Consistency checks:}\;
    \quad $s_{n,\mathcal{I}\pm} + f_{n,\mathcal{I}\pm} = P_{n,\mathcal{I}}$\;
    \quad $s_{n\pm} + f_{n\pm} = 1$\;
\end{algorithm}

While the above-mentioned procedure works in principle, the calculations scale exponentially with $n$ and take increasingly longer time. The difficulty arises from the fact that the coefficients for the various terms in the polynomial (\ref{polynomial-form}) have many contributions, so adding over them takes a lot of operations. One possible way to simplify this process is to set any creation operators for which there are no photons in a given click pattern to be zero. Then the polynomial simplifies a bit and the coefficients require somewhat fewer operations, though the overall calculation still scales exponentially. In our work, we have employed this approach and this has allowed us to calculate a few more results than what would otherwise have been possible.

\subsection{Results broken down in terms of total photon counts $\scriptI$}
\label{results-broken-down-in-terms-photon-counts}

Here are the patterns we observe based on the break-down of our results for the success and failure rates in terms of the total number of photons in the various output click patterns.
\begin{enumerate}
\item Click patterns with 0 or $n$ photons in the output give zero success in measuring $|+_\phi\ra$ or $|-_\phi\ra$. That is, $s_{n, 0\pm} = s_{n, n\pm} = 0$.
\item We get a 0 success rate for measuring $|+_\phi\ra$ except when we have $\scriptI={n}/{2}$ photons in the output click patterns. Thus, for odd $n$, for which an integer ${n}/{2}$ does not exist, the total success rate for $|+_\phi\ra$ is zero.
\item For ${n}/{2}$ photons in the output click patterns, we get perfect discrimination between $|+_\phi\ra$ and $|-_\phi\ra$ with the success probabilities
\be
s_{n, \frac{n}{2}\pm} 
= P_{n,\frac{n}{2}}
= \frac{1}{2^n}\binom{n}{\frac{n}{2}},
\label{success-rate-n-over-2}
\ee
where in $s_{n, {n}/{2}\pm}$, we are using the notation $s_{n, \scriptI\pm}$, in which $\scriptI$ is the total number of photons in the output click patterns being considered which in this case is $\scriptI= {n}/{2}$.
Note that this decreases as $n$ gets larger and goes to $0$ as $n\to\infty$. This explains why the success rate for $|+_\phi\ra$ goes to zero in the large $n$ limit.
\item The total success probability for measuring $|-_\phi\ra$ from click patterns with a total of $\scriptI$ photons is
\be
s_{n,\scriptI-} = \frac{4(n-\scriptI)\scriptI \binom{n}{\scriptI}}{2^n n^2} 
\label{success-rate-n-i-minus}
\ee
Although we do not have a formal proof for this formula for all $n$ and $\scriptI$, we describe how we obtained it in the next section.
We should mention here that it is a simple exercise to see that this reduces to (\ref{success-rate-n-over-2}) when we take $\scriptI={n}/{2}$.
\end{enumerate}
While we do not have a rigorous proof for any of the above expressions, we show in the next section how we can obtain them.

\subsection{Expressions for the total success rates for $\scriptI$ photon click patterns}
\label{success-rates-different-i}

We will now show how we obtain the expressions for the success rates $s_{n, \scriptI\pm}$ that we observe and report in the previous section. For convenience, we will work this out for the $\phi=0$ case, so that $|\pm_\phi\ra = |\pm\ra = [|0\ra\pm |1\ra]/\sqrt{2}$, but our results will be applicable for all $\phi$ as we have proved in the main text.

Let $|+_{\scriptI}\ra$ and $|-_{\scriptI}\ra$ be the (normalized) projections of $|+_1 +_2\ldots +_n\ra$ and $|-_1+_2\ldots +_n\ra$, respectively, on to the subspace of states where $\scriptI$ qubits are in $|1\ra$, and the rest are in $|0\ra$:
\begin{align}
|+_{\scriptI}\ra = \frac{1}{\sqrt{\binom{n}{\scriptI}}} \sum_{\substack{i_1,i_2,\ldots ,i_n=0,\\ i_1+i_2+\ldots +i_n=\scriptI}}^1 |i_1 i_2 i_3 \ldots i_n\ra \nn \\
|-_{\scriptI}\ra = \frac{1}{\sqrt{\binom{n}{\scriptI}}} \sum_{\substack{i_1,i_2,\ldots ,i_n=0,\\ i_1+i_2+\ldots +i_n=\scriptI}}^1 (-1)^{i_1} |i_1 i_2 i_3 \ldots i_n\ra
\label{plus-and-minus-i-states-definitions}
\end{align}
Let the unitary transformation associated with our linear interferometer be $U_{\rm lin}$, so that the $|+_{\scriptI}\ra$ and $|-_{\scriptI}\ra$ states going into the interferometer give the output states
\begin{align}
|+_{\scriptI} ^\prime\ra &= U_{\rm lin} |+_{\scriptI}\ra \nn \\
|-_{\scriptI} ^\prime\ra &= U_{\rm lin} |-_{\scriptI}\ra
\label{plus-and-minus_{scriptI}-prime-definitions}
\end{align}
To measure our single-rail qubit in the $X$ basis, we need to discriminate between the states $|+_{\scriptI} ^\prime\ra$ and $|-_{\scriptI} ^\prime\ra$ for each $\scriptI$.
Now, consider the case when these have the forms
\begin{align}
|+_{\scriptI} ^\prime\ra &= \sum_{\substack{i_1, i_2,\ldots ,i_n=0,\\ i_1+i_2+\ldots +i_n=\scriptI}}^{\scriptI} d_{+, i_1 i_2\ldots i_n} |i_1i_2\ldots i_n\ra \nn \\
|-_{\scriptI} ^\prime\ra &= \sum_{\substack{i_1, i_2,\ldots ,i_n=0, \\i_1+i_2+\ldots +i_n=\scriptI}}^{\scriptI} \left(\gamma_{n,\scriptI} d_{+, i_1 i_2\ldots i_n} \,+\,
d_{-, i_1 i_2\ldots i_n}\right) |i_1i_2\ldots i_n\ra
\label{gamma-form-i}
\end{align}
with $0\leq |\gamma_{n,\scriptI}| \leq 1$,
and $d_{+, i_1 i_2\ldots i_n} d_{-, i_1 i_2\ldots i_n} = 0$ for any given $|i_1 i_2 \ldots i_n\rangle$ configuration \ie, one of $d_{+, i_1 i_2\ldots i_n}$ and $d_{-, i_1 i_2\ldots i_n}$ is zero. This way, when $\gamma_{n,\scriptI} \neq 0$,
the output photon click patterns $|i_1i_2\ldots i_n\ra$
in $|+_{\scriptI} ^\prime\ra$
also appear in $|-_{\scriptI} ^\prime\ra$,
whereas $|-_{\scriptI} ^\prime\ra$ also has some click patterns not in $|+_{\scriptI} ^\prime\ra$.
But when $\gamma_{n,\scriptI} =0$, then $|+_{\scriptI} ^\prime\ra$
and $|-_{\scriptI} ^\prime\ra$ give  totally non-overlapping click patterns, representing perfect distinguishability between the single-rail qubit being found to be in $|+\rangle$ and $|-\rangle$ when the total number of photons in the obtained photon click pattern is $\scriptI$.

Now, if $|\pm_{\scriptI} ^\prime\ra$ have the forms (\ref{gamma-form-i}), then given that we have a total of $\scriptI$ photons in an output click pattern, the conditional success probability for correctly measuring our single-rail qubit to be in $|+\ra$ will be 0 when $\gamma_{n,\scriptI}\neq 0$, and 1 when $\gamma_{n,\scriptI}=0$.
And for measuring the single-rail qubit to be in $|-\ra$, we will have a conditional failure probability of $\gamma_{n,\scriptI} ^2$, which will translate into a success rate of $1-\gamma_{n,\scriptI} ^2$.
Multiplying these conditional probabilities by the probability $P_{n,\scriptI}$ that we have a click pattern with $i$ total photons then gives the overall success rates for correctly measuring our single-rail qubit to be in $|+\ra$ and $|-\ra$:
\begin{align}
s_{n+} &= 0 \,\, \text{ for }\, \gamma_{n,\scriptI} \neq 0 \nn \\
s_{n+} &= P_{n,\scriptI} \,\, \text{ for }\, \gamma_{n,\scriptI} =0 \nn \\
s_{n-} &= P_{n,\scriptI} \left(1-\gamma_{n,\scriptI} ^2\right),
\label{success-failure-rates-i-gamma}
\end{align}
with $\gamma_{n,\scriptI} = \la +_{\scriptI} ^\prime|-_{\scriptI} ^\prime\ra = \la +_{\scriptI}|-_{\scriptI}\ra$, where we have used the definitions (\ref{plus-and-minus_{scriptI}-prime-definitions}).

It is also straightforward to argue the converse of the above: if the success and failure rates have the form \eqref{success-failure-rates-i-gamma}, then it means the states $|+_{\scriptI} ^\prime\ra$ and $|-_{\scriptI} ^\prime\ra$ have the forms~\eqref{gamma-form-i}. It turns out that our success and failure rates for $|+\ra$ and $|-\ra$ for the power-of-2 Hadamard codes and QFTs that we have worked out, indeed have the above forms. This can for instance be seen clearly in the 2- and 4-component Hadamard code cases explicitly worked out in the next two sections.

What remains to be done is to calculate $\gamma_{n,\scriptI}$. 
Recalling (\ref{plus-and-minus-i-states-definitions}), note that
$|+_{\scriptI}\ra$ has a total of $\binom{n}{\scriptI}$ terms, all of which have plus signs.
On the other hand, $|-_{\scriptI}\ra$ has a total of $\binom{n}{\scriptI}$ terms, out of which $\binom{n-1}{\scriptI-1}$ terms have minus signs, and the remaining
$\binom{n}{\scriptI} -\binom{n-1}{\scriptI-1}$ terms have positive signs.
The inner product $\la +_{\scriptI}|-_{\scriptI}\ra$ is the number of positive terms minus the number of negative terms, up to the overall ${1}/{\binom{n}{\scriptI}}$ factor arising from the normalization factors in $|+_{\scriptI}\ra$ and $|-_{\scriptI}\ra$.
We thus obtain
\begin{align}
\gamma_{n,\scriptI} &= \la +_{\scriptI}|-_{\scriptI}\ra \nn \\
&= \frac{1}{\binom{n}{\scriptI}} \left[\binom{n}{\scriptI} -2\binom{n-1}{\scriptI-1}\right] \nn \\
&= \frac{n-2\scriptI}{n}
\end{align}
Note that this is zero when $\scriptI={n}/{2}$, giving us perfect discrimination between $|+\ra$ and $|-\ra$.
Inserting this into (\ref{success-failure-rates-i-gamma})
along with $P_{n,\scriptI} = {\binom{n}{\scriptI}}/{2^n}$
from (\ref{P_i-formula}), we then obtain the success rates
\begin{align}
s_{n, \scriptI+} &= 0 \,\, \text{ for }\, \scriptI\neq\frac{n}{2} \nn \\
s_{n, \scriptI+} &= \frac{\binom{n}{\scriptI}}{2^n} \,\, \text{ for }\, \scriptI=\frac{n}{2} \nn \\
s_{n, \scriptI -} &= \frac{4(n-\scriptI)\scriptI \binom{n}{\scriptI}}{2^n n^2} 
\label{success-failure-rates-i-gamma}
\end{align}
This is the expression we reported in (\ref{success-rate-n-i-minus}) in the previous section.

\subsection{The 2-component Green Machine, \ie~a 50-50 beam splitter}

Here we describe the explicit analytical calculations of the $n=2$ case for an $X$ basis measurement where we use $|+\ra$ ancillas. Taking $\phi=0$, recall that the input states $|\pm_1 +_2\ra$ are created by the operators
\begin{align}
\frac{1}{2} (1+a_1 ^\dagger) (1+a_2 ^\dagger)
&= \frac{1}{2} (1+a_1 ^\dagger +a_2 ^\dagger +a_1 ^\dagger a_2 ^\dagger)|0_1 0_2\ra \nn \\
\frac{1}{2} (1+a_1 ^\dagger) (1+a_2 ^\dagger) &= \frac{1}{2} (1-a_1 ^\dagger +a_2 ^\dagger -a_1 ^\dagger a_2 ^\dagger)
\end{align}
Now, consider feeding these into a 50-50 beam splitter which replaces $a_1 ^\dagger$ and $a_2 ^\dagger$ by the output creation operators
\begin{align}
a_1 ^\dagger &\mapsto (b_1 ^\dagger +b_2 ^\dagger)/\sqrt{2} \nn \\
a_2 ^\dagger &\mapsto (b_1 ^\dagger -b_2 ^\dagger)/\sqrt{2}
\end{align}
After a bit of simplification, we obtain
\begin{align}
\frac{1}{2}(1+a_1 ^\dagger)(1+a_2 ^\dagger) 
&\mapsto \frac{1}{2}\left(1+\sqrt{2}b_1 ^\dagger +[(b_1 ^\dagger)^2 -(b_2 ^\dagger)^2]\right) \nn \\
\frac{1}{2}(1-a_1 ^\dagger)(1+a_2 ^\dagger) 
&\mapsto \frac{1}{2}\left(1-\sqrt{2}b_2 ^\dagger -[(b_1 ^\dagger)^2 -(b_2 ^\dagger)^2]\right)
\end{align}
These tell us that we obtain the click patterns
$|1_1 0_2\ra$ and $|0_1 1_2\ra$
from the input states $|+_1+_2\ra$ and $|-_1+_2\ra$, respectively, with probability $1/2$.
These click patterns give $50\%$ success rate. The zero or two-photon click patterns correspond to failure of measuring our single-rail qubit in the $X$ basis.

\subsection{The 4-component Green Machine}

We now show the explicit analytical calculations of the $n=4$ case for an $X$ basis measurement using $|+\ra$ ancillas and a 4-component Green Machine. We again start with the operators that act on the vacuum to generate the $|\pm_1 +_2 +_3 +_4\ra$ input states:
\be
\frac{1}{4}
(1\pm a_1 ^\dagger) (1+a_2 ^\dagger) (1+a_3 ^\dagger) (1+a_4 ^\dagger)
\ee
We then apply the first beam splitter layer of the 4-component Green Machine according to the recursive construction described earlier. In this layer, one beam splitter mixes ports 1 and 2 with the transformations
\begin{align}
a_1 ^\dagger &\mapsto \frac{1}{\sqrt{2}}(c_1+c_2), \nn \\
a_2 ^\dagger &\mapsto \frac{1}{\sqrt{2}}(c_1-c_2),
\end{align}
and the second beam splitter mixes ports 3 and 4 with the transformations
\begin{align}
a_3 ^\dagger &\mapsto \frac{1}{\sqrt{2}}(c_3+c_4), \nn \\
a_4 ^\dagger &\mapsto \frac{1}{\sqrt{2}}(c_3-c_4).
\end{align}
Here, the $c_i$ operators are creation operators for ports number $i=1,\ldots, 4$, and we have taken the liberty not to put daggers on them in order to simplify the notation. We can afford to take this liberty since we do not have any annihilation operators anywhere in this entire analysis, and will therefore take the daggers to be implicit in the rest of this section.
 
Now, for the input state $|+_1 +_2 +_3 +_4\ra$, we get
\be
\frac{1}{4}
[1+ \sqrt{2} b_1 +(b_1 ^2 -b_2 ^2)/2]
[1+ \sqrt{2} b_3 +(b_3 ^2 -b_4 ^2)/2]
\ee
For the input state $|-_1 +_2 +_3 +_4\ra$, we get
\be
\frac{1}{4}
[1-\sqrt{2} b_2 -(b_1 ^2 -b_2 ^2)/2]
[1+ \sqrt{2} b_3 +(b_3 ^2 -b_4 ^2)/2] 
\ee
Now, apply the second layer of beam splitters. One beam splitter mixes ports 1 and 3, and another one mixes ports 2 and 4. The transformations are
\begin{align}
c_1 &\mapsto \frac{1}{\sqrt{2}}(d_1+d_3), \nn \\
c_3 &\mapsto \frac{1}{\sqrt{2}}(d_1-d_3)
\end{align}
for one beam splitter, and
\begin{align}
c_2 &\mapsto \frac{1}{\sqrt{2}}(d_2+d_4), \nn \\
c_4 &\mapsto \frac{1}{\sqrt{2}}(d_2-d_4)
\end{align}
for the other beam splitter, with $d_j$ the photon creation operators for the output of the 4-component Green Machine for $j\in \{1, \ldots ,4\}$.

For the input state $|+_1 +_2 +_3 +_4\ra$ going into our 4-component Green Machine, we get

\begin{align}
& \frac{1}{4}
\left[1+ (d_1+d_3) +\frac{1}{4}((d_1+d_3)^2 -(d_2+d_4)^2)\right]
\left[1+ (d_1-d_3) +\frac{1}{4}((d_1-d_3)^2 -(d_2-d_4)^2)\right] \nn \\
& =\frac{1}{4}
\left[1+2d_1  +\frac{1}{2}(3d_1 ^2 -d_2 ^2 -d_3 ^2 -d_4 ^2) + \frac{1}{4}(d_1-d_3) [(d_1+d_3)^2 -(d_2+d_4)^2]
+\frac{1}{4}(d_1+d_3) [(d_1-d_3)^2 -(d_2-d_4)^2] \right.\nn \\
&\left.+\frac{1}{16}[(d_1+d_3)^2 -(d_2+d_4)^2] [(d_1-d_3)^2 -(d_2-d_4)^2]\right].
\label{plus-4-compponent-final-polynomial}
\end{align}
For $|-_1 +_2+_3+_4\ra$, we get
\begin{align}
&\frac{1}{4}
\left[1- (d_2+d_4) - \frac{1}{4}((d_1+d_3)^2 -(d_2+d_4)^2)\right]
\left[1+ (d_1-d_3) + \frac{1}{4}((d_1-d_3)^2 -(d_2-d_4)^2)\right] \nn \\
&= \frac{1}{4}
\left[1+d_1 -d_2-d_3-d_4 -d_1(d_2+d_3+d_4)
+d_2d_3 +d_2d_4 +d_3d_4  - \frac{1}{4}(d_1-d_3) [((d_1+d_3)^2 -(d_2+d_4)^2] \right. \nn \\
&\left.
- \frac{1}{4}(d_2+d_4) [(d_1-d_3)^2 -(d_2-d_4)^2] -\frac{1}{16}[(d_1+d_3)^2 -(d_2+d_4)^2] [(d_1-d_3)^2 -(d_2-d_4)^2]\right].
\label{minus-4-compponent-final-polynomial}
\end{align}

Now, we compare (\ref{plus-4-compponent-final-polynomial})
and (\ref{minus-4-compponent-final-polynomial}) line by line.
In the linear terms, we see that for $|+\ra$ we have $d_1/2$, and $d_1/4$ also appears in the $|-\ra$ case.
Thus for 1-photon states, we have total failure for $|+\ra$, and a failure rate of $({1}/{4})^2 = {1}/{16}$ for $|-\ra$.
Since the probability for obtaining one of the click patterns with 1- total photon click is $P_{4,1} = {\binom{4}{1}}/{2^4} = {1}/{4}$, we obtain a success rate of
\be
s_{4, 1-} = \frac{1}{4} - \frac{1}{16} = \frac{3}{16}
\ee
for correctly measuring $|-\ra$ from 1-photon click patterns, which matches the formula (\ref{success-rate-n-i-minus}) we gave while describing the general pattern in terms of the photon counts in the various click patterns.

For click patterns with 2 photons, we compare the quadratic terms. Note that there are no common quadratic terms in (\ref{plus-4-compponent-final-polynomial}) and (\ref{minus-4-compponent-final-polynomial}). This means that we have full success from 2-photon click patterns:
\be 
s_{4, 2\pm} = P_{4,2} = \frac{\binom{4}{2}}{2^4} = \frac{3}{8}
\ee
Again, this matches the results (\ref{success-rate-n-over-2}) while describing the general patterns in terms of the break down in terms of the total photon counts in the output click patterns.

Next, we come to the qubit terms. With a bit of simplification, the qubit terms in (\ref{plus-4-compponent-final-polynomial}) simplify to
\be
\frac{1}{8}d_1 ^3
-\frac{1}{8}d_1 (d_2 ^2 +d_3 ^2 +d_4 ^2)
+\frac{1}{4}d_2 d_3 d_4
\label{4-component-qubit-plus}
\ee
In the qubit terms in (\ref{minus-4-compponent-final-polynomial}),
it is clear that the $(d_2+d_4) [(d_1-d_3)^2 -(d_2-d_4)^2]/16$
part will have nothing overlapping with this.
However, if we simplify
$- (d_1-d_3) [(d_1+d_3)^2 -(d_2+d_4)^2]/16$,
then we find that it has the part
\be
-\frac{1}{16}d_1 ^3
+\frac{1}{16}d_1 (d_2 ^2 +d_3 ^2 +d_4 ^2)
-\frac{1}{8} d_2 d_3 d_4
\label{4-component-qubit-minus-common}
\ee
plus some additional uncommon terms.
Thus, everything in the qubit part for the $|+\ra$ state for our single-rail qubit is also appearing in the output state corresponding to $|-\ra$, and hence we have zero probability of success for correctly measuring $|+\ra$ from 3-photon click patterns.
For $|-\ra$, we have half of the polynomial that appears in $|+\ra$, so the failure probability is
${P_{4,3}}/{4}$, with the factor of ${1}/{4}$ arising from squaring ${1}/{2}$.
Thus the total success rate associated with 3-photon click patterns is
 \be
s_{4,3-} = P_{4,3} -\frac{P_{4,3}}{4} = \frac{3}{16}
\ee
Again, this is consistent with the formula (\ref{success-rate-n-i-minus}) from the results we obtained in terms of the photon counts.

Combining these results, the overall success probabilities are
\begin{align}
s_{4+} &= 0+\frac{3}{8}+0 = \frac{3}{8} \nn \\
s_{4-} &= \frac{3}{16} +\frac{3}{8}+\frac{3}{16} = \frac{3}{4} \nn \\
s_{4, \rm overall} &= \frac{1}{2}\left(s_{4+} +s_{4-}\right) = \frac{9}{16}
\end{align}

\subsection{Non-power of 2 Hadamard unitaries and results for the 12-component case}

While power of 2 Hadamard unitaries have the above-mentioned beautiful and recursive form, Hadamard matrices have also been shown to exist for several numbers other than powers of 2 with the general conjecture that these exist for all multiples of 4~\cite{Paley1933}.
For example, one construction for $n=12$ gives the unitary transformation matrix
\be
U_{\mathrm{GM}, 12} \equiv \frac{1}{\sqrt{12}}\left(
\begin{array}{cccccccccccc}
 1 & 1 & 1 & 1 & 1 & 1 & 1 & 1 & 1 & 1 & 1 & 1 \\
 1 & -1 & 1 & -1 & 1 & 1 & 1 & -1 & -1 & -1 & 1 & -1 \\
 1 & -1 & -1 & 1 & -1 & 1 & 1 & 1 & -1 & -1 & -1 & 1 \\
 1 & 1 & -1 & -1 & 1 & -1 & 1 & 1 & 1 & -1 & -1 & -1 \\
 1 & -1 & 1 & -1 & -1 & 1 & -1 & 1 & 1 & 1 & -1 & -1 \\
 1 & -1 & -1 & 1 & -1 & -1 & 1 & -1 & 1 & 1 & 1 & -1 \\
 1 & -1 & -1 & -1 & 1 & -1 & -1 & 1 & -1 & 1 & 1 & 1 \\
 1 & 1 & -1 & -1 & -1 & 1 & -1 & -1 & 1 & -1 & 1 & 1 \\
 1 & 1 & 1 & -1 & -1 & -1 & 1 & -1 & -1 & 1 & -1 & 1 \\
 1 & 1 & 1 & 1 & -1 & -1 & -1 & 1 & -1 & -1 & 1 & -1 \\
 1 & -1 & 1 & 1 & 1 & -1 & -1 & -1 & 1 & -1 & -1 & 1 \\
 1 & 1 & -1 & 1 & 1 & 1 & -1 & -1 & -1 & 1 & -1 & -1 \\
\end{array}
\right).
\label{12-component-GM-unitary}
\ee

Note that this is not a symmetric matrix and its transpose will also give an equally valid normalized Hadamard matrix. Thus, non power-of-2 Hadamard matrices do not have the nice symmetries that the power-of-2 cases have. Nevertheless, as an example of the non-power of 2 case, we have calculated the success and failure rates for the above-mentioned 12-component unitary transformation. We find that this case substantially deviates from the QFT and the power of 2 Hadamard codes and gives a much lower overall success rate of about $36.63\%$. This deviation is not entirely surprising considering that, unlike power-of-2 Hadamard codes or QFTs, the 12-component Hadamard matrix is not so nice and symmetric.

For details, the overall success and failure rates for measuring $|+\ra$ and $|-\ra$ are
\begin{align}
s_{12+} &= \frac{6731395}{47775744} \approx 0.140956604 \nn \\
s_{12-} &= \frac{6106045627}{10319560704} \approx 0.5916962749 \nn \\
f_{12+} &= \frac{41044349}{47775744} \approx 0.8591043396 \nn \\
f_{12-} &= \frac{4213515077}{10319560704} \approx 0.4083037251
\end{align}
The break-up in terms of the total photons $i$ in the output click-patterns is given in table~\ref{table:12component_results}.
\begin{table}[h!]
\centering 
\caption{Values of \( s_{12, \scriptI+} \) and \( s_{12, \scriptI-} \)}
\[
\begin{array}{ccc}
\toprule
\text{No. of photons } \scriptI & s_{12, \scriptI+} & s_{12, \scriptI-} \\
\midrule
0 & 0 & 0 \\
1 & 0 & {11}/{12288} \\
2 & 0 & {55}/{6144} \\
3 & 0 & {1375}/{36864} \\
4 & 0 & {605}/{6912} \\
5 & 0 & {7535}/{55296} \\
6 & {121385}/{884736} & {16093}/{110592} \\
7 & 0 & {210595}/{1990656} \\
8 & 0 & {3685}/{73728} \\
9 & {774455}/{214990848} & {2291245}/{143327232} \\
10 & {9295}/{143327232} & {1817585}/{573308928} \\
11 & {6325}/{214990848} & {3182113}/{10319560704} \\
12 & 0 & 0 \\
\bottomrule
\end{array}
\]
\label{table:12component_results}
\end{table}

\subsection{Using coherent state ancillas instead of $|+\rangle$}

As before, let us label the optical mode encoding our single-rail qubit, which we wish to measure in the $X$ basis, as mode 1, and consider mixing it with a coherent state ancilla $|\alpha\rangle$ in mode 2, using a fifty-fifty beam splitter of the form (\ref{beam-splitter-convention})~\footnote{The idea of employing a coherent state ancilla and mixing it with the optical mode encoding a single-rail qubit, was described in~\cite{Khabiboulline2019, Stas2026} in the context of loading the joint state of a star light photon arriving across two distant telescopes on to quantum memories at the two sites. In that set up, the proposal was to employ a coherent state $|\alpha\rangle$ at each site, mix it with the optical mode of the star light photon at that location using a balanced beam splitter, and measure the outputs with photon number detectors. Then Alice and Bob share their measurement results with each other through classical communication to determine if the two-telescope state of the astronomical photon has been correctly loaded on to quantum memories at the two sites, loaded with a minus sign that needs to be corrected with a $Z$ gate, or if there is a failure in loading. Here, we consider the problem of making an $X$ basis measurement on just one single-rail qubit, or loading its state on to a quantum memory by employing a coherent state ancilla as opposed to the above-mentioned two-site scenario.}. We will denote the coherent state in mode 2 as $|\alpha_2\rangle$ with the subscript 2 indicating the mode number.
Note that if we mix the coherent state $|\alpha_2\rangle$ with the Fock states $|0\rangle$ or $|1\rangle$ in the first mode, we obtain
\begin{align}
|0_1 \alpha_2\rangle
&\to \left|\left(\alpha/\sqrt{2}\right)_1, \left(-\alpha/\sqrt{2}\right)_2\right\rangle \nn \\
|1_1 \alpha_2\rangle
&\to \frac{1}{\sqrt{2}}(b_1 ^\dagger + b_2 ^\dagger) \left|\left(\alpha/\sqrt{2}\right)_1, \left(-\alpha/\sqrt{2}\right)_2\right\rangle
\end{align}
Here, $0_1$ and $1_1$ on the LHS denote the 0 and 1-photon Fock states in input mode 1, and $b_1 ^\dagger$ and $b_2 ^\dagger$ are the creation operators for the two output modes of the beam splitter, which we still label as modes 1 and 2.

To successfully measure the input state of mode 1 in the $|\pm\rangle$ basis, we need both the above terms to have the same total number of photons in order to scramble the information about whether we had an incoming photon in mode 1 or not.
 Putting the above two equations together as
$(|0_1 \alpha_2\rangle \pm |1_1 \alpha_2\rangle)/\sqrt{2}$,
and expanding the coherent states in terms of the creation operators and Fock states, this condition gives the terms
\be
e^{-|\alpha|^2/2} \frac{(-1)^j}{\sqrt{2}} 
\left[\frac{1}{i! j!} (\alpha/\sqrt{2})^{i+j} 
\pm \frac{(-1)^j}{\sqrt{2}}\left(\frac{1}{(i-1)! j!}
-\frac{1}{i! (j-1)!}\right)
(\alpha/\sqrt{2})^{i+j-1}
\right]
(b_1 ^\dagger)^i (b_2 ^\dagger)^j |0_1 0_2\rangle
\ee
\be
= e^{-|\alpha|^2/2} 
\frac{(-1)^j}{\sqrt{2 i! j!}} (\alpha/\sqrt{2})^{i+j}
\left[1\pm (i-j)/\alpha\right]
|i_1 j_2\rangle
\label{coherent-state-ancilla-post-beam-splitter-state}
\ee
This gives the condition
$\alpha = \pm (i-j)$
for a successful $X$ basis measurement of mode 1.
Assuming positive values of $\alpha$, the click patterns $|i, i-\alpha\rangle$ and $|i-\alpha, i\rangle$ mean the signal-rail qubit being successfully measured and found to be in $|+\rangle$ and $|-\rangle$, respectively, both with the probability
\be
P_{s, 2, i, \alpha} = \frac{2e^{-\alpha^2}}{i! (i-\alpha)!}
(|\alpha|^2/2)^{2i-\alpha},
\,\, |\alpha| =1, 2,\ldots,
\, \mathrm{and} \, i=|\alpha|, |\alpha|+1, |\alpha|+2, \ldots 
\ee
where the subscript $s$ in $P_{s, 2, i, \alpha}$ denotes the fact that this is the probability of success, and 2 indicates that we are applying a balanced beam splitter, i.e., a 2-component Green Machine.
The total probability of success is thus
\be
P_{s, 2, \alpha} = \sum_{i=\alpha}^\infty P_{s, 2, i,\alpha}
= 2e^{-\alpha^2}\, I_{|\alpha|}\!\left(\alpha^2\right),
\,\, \alpha=\pm 1,\pm 2,\ldots
\ee
where $I_{|\alpha|}\!\left(\alpha^2\right)$ is the modified Bessel function of the first kind of order $|\alpha|$ with argument $\alpha^2$.
This gives a maximum success rate of approximately $0.4158$ for $|\alpha| =1$ and decreasing success rates with increasing $|\alpha|$.
Note that replacing $\alpha$ with $-\alpha$ gives the same success rates, with the only change being that the click patterns associated with the $|+\rangle$ and $|-\rangle$ outcomes are switched.

This method of course requires a fine-tuned value of $\alpha$, and is therefore prone to error. In the context of loading a single-rail qubit in some arbitrary state $|\psi_1\rangle \equiv \upsilon |0_1\rangle +\xi |1_1\rangle$
onto a quantum memory, we can ask what actual state is loaded onto the latter if $\alpha$ is not perfectly fine-tuned to the correct value.
Recalling the state transfer protocol from the motivation section \ref{motivation-section} of the supplementary material,
we initialize the quantum memory in $|\overline{0}_a\rangle$ and apply a CNOT gate from our single-rail qubit to this memory qubit. But now use a coherent state $|\alpha\rangle$ in mode 2.
The joint state of the two optical modes and the quantum memory after the CNOT operation but before the beam splitter is
\be
(\upsilon + \xi a_1 ^\dagger X_a) |0_1 \alpha_2 \overline{0}_a\rangle,
\ee
where $X_a$ is a Pauli $X$ operator on the quantum memory.
Recall that we need the coherent state to provide one additional photon when mixing with $|0_1\rangle$ than it does when mixing with $|1_1\rangle$ for a successful $X$ basis measurement.
Applying this condition and repeating the steps involved in deriving equation (\ref{coherent-state-ancilla-post-beam-splitter-state}), we find that the state after mixing the two optical modes in a balanced beam splitter is
\be
(-1)^j e^{-|\alpha|^2/2}  
\frac{1}{\sqrt{i! j!}} (\alpha/\sqrt{2})^{i+j}
\left[\alpha \upsilon + (i-j) \xi X_a\right]/\alpha
|i_1 j_2 \overline{0}_a\rangle
\ee
\be
= \frac{(-1)^j e^{-|\alpha|^2/2}}{\alpha} 
\frac{1}{\sqrt{i! j!}} (\alpha/\sqrt{2})^{i+j}
\sqrt{|\alpha\upsilon|^2 +(i-j)^2 |\xi|^2}
\frac{\alpha \upsilon + (i-j) \xi X_a}{\sqrt{|\alpha\upsilon|^2 + (i-j)^2|\xi|^2}}
|i_1 j_2 \overline{0}_a\rangle
\label{coherent-state-ancilla2},
\ee
where we are using the same numbers to label the input and output modes of the beam splitter.
Thus, if we obtain the click pattern $|i_1 j_2\rangle$ in a photon number detection of the beam splitter outputs, the quantum memory ends up in the state
\be
|\psi_{\rm {loaded}}\rangle
\equiv \frac{1}{\sqrt{|\alpha\upsilon|^2 +(i-j)^2|\xi|^2}}
[\alpha \upsilon |\overline{0}_a\rangle
+ (i-j) \xi|\overline{1}_a\rangle]
\ee
with the probability
\be
P_{\rm {loading}, \alpha, i,j}
= \frac{e^{-|\alpha|^2}}{|\alpha|^2} 
\frac{1}{i! j!} (|\alpha|^2/2)^{i+j}
\left[|\alpha\upsilon|^2 +(i-j)^2|\xi|^2\right].
\ee
Note that the click pattern with the values of $i$ and $j$ interchanged, has the same probability and gives the same state as $|\psi_{\rm {loaded}}\rangle$, but with a `$-$' sign on the $|\overline{1}_a\rangle$ term. Since this sign can be flipped with a $Z$ gate, we thus again obtain $|\psi_{\rm loaded}\rangle$ with the same probability.
The combined probability of obtaining $|\psi_{\rm loaded}\rangle$ on obtaining the click patterns $|i_1 j_2\rangle$ and $|j_1 i_2\rangle$  is then
$2P_{\rm loading, \alpha, i,j}$. And the total probability of this is the sum over all values of $i$ and $j$ for a fixed $|i-j|$.
Since the highest success rate for a perfect $X$ basis measurement of the single rail qubit was obtained for $|i-j|=1$ with $\alpha=\pm 1$, we interpret the click patterns $|i_1 (i-1)_2\rangle$ and $|(i-1)_1 i_2\rangle$ as representing successful state loading, albeit with possible imperfection due to the value of $\alpha$ not being perfectly fine-tuned. The total success probability of loading the single-rail qubit on to a quantum memory is then
\be
P_{\rm {tot}, \alpha}
\equiv \sum_{i=1}^\infty 2P_{\rm {loading}, \alpha, i,i-1}
= \frac{2e^{-|\alpha|^2}}{|\alpha|^2} 
\sum_{i=1}^\infty \frac{1}{i! (i-1)!} (|\alpha|^2/2)^{2i-1}
[|\alpha\upsilon|^2 +|\xi|^2].
\ee

Next, let us consider employing 3 coherent state ancillas $|\alpha\rangle$ and mixing them with our single-rail qubit in a 4-component Green Machine to perform an $X$ basis measurement. We feed the single-rail qubit into the first input, and the ancillas into the rest. Recall from the recursive construction of a Green Machine described in section \ref{Green-machine-construction}, that a 4-component Green Machine first involves two separate balanced beam splitters of the form (\ref{beam-splitter-convention}) mixing mode 1 with 2, and 3 with 4, and in step two, we mix mode 1 with 3, and 2 with 4 using similar 50-50 beam splitters. Consequently, we obtain
\begin{align}
|0_1 \alpha_2 \alpha_3 \alpha_4\rangle
&\to |\left(3\alpha/2\right)_1, \left(-\alpha/2\right)_2, \left(-\alpha/2\right)_3, \left(-\alpha/2\right)_4\rangle, \nn \\
|1_1 \alpha_2 \alpha_3 \alpha_4\rangle
&\to \frac{1}{2}(b_1 ^\dagger + b_2 ^\dagger +b_3 ^\dagger + b_4 ^\dagger) |\left(3\alpha/2\right)_1, \left(-\alpha/2\right)_2, \left(-\alpha/2\right)_3, \left(-\alpha/2\right)_4\rangle.
\end{align}
Here, as in the case of a single coherent state ancilla, $0_1$ and $1_1$ denote the 0 and 1-photon Fock states in the first input mode of the Green Machine, and $b_1 ^\dagger$, $b_2 ^\dagger$, $b_3 ^\dagger$ and $b_4 ^\dagger$ are the creation operators for the output modes of the 4-component Green Machine. In the second equation, we have used the fact that the creation operator $a_1 ^\dagger$ for a photon going into input mode 1 goes to
$\frac{b_1 ^\dagger + b_2 ^\dagger + b_3 ^\dagger + b_4 ^\dagger}{2}$ under the action of the Green Machine.

Now, again, the coherent states need to provide 1 additional photon to mix with the $|0_1\rangle$ state compared to $|1_1\rangle$ to scramble the information about whether we have $|0_1\rangle$ or $|1_1\rangle$ and provide a successful $X$ basis measurement.
Putting together the above two equations as $|\pm_1 \alpha_2 \alpha_3 \alpha_4\rangle$
and applying this condition now gives the terms in the output state 
\be
\exp(-3|\alpha|^2/2)
\frac{1}{\sqrt{2 i! j! k! l!}}
3^{i} (\alpha/2)^{i +j+k+l}
\left[1 \pm (i/3 -j -k-l)/\alpha\right]
|i_1 j_2 k_3 l_4\rangle,
\ee
where $|i_1 j_2 k_3 l_4\rangle$ means the output photon click pattern with $i$, $j$, $k$ and $l$ photons in modes 1, 2, 3 and 4, respectively. 
We thus have the conditions
$\alpha = i/3 -j -k -l$
and
$\alpha = -(i/3 -j -k -l)$
for a successful measurement in $|+_1\rangle$
and $|-_1\rangle$, respectively.

The success rates are thus
\be
P_{s\pm, 4, \alpha}
= 2\exp(-3|\alpha|^2)
\sum_{i, j, k, l=0, i/3-j-k-l =\pm \alpha}^\infty
\frac{9^i}{i! j! k! l!}
(|\alpha|^2/4)^{i +j+k+l}
\label{4-component-GM-coherent-state-success-rate}.
\ee 
Here, the $\pm$ in the subscript of $P_{s\pm, 4, \alpha}$ indicates whether it is the success rate for measuring $|+\rangle$ or $|-\rangle$, and 4 and $\alpha$ indicate that this is for a 4-component Green Machine with coherent states $|\alpha\rangle$ used as the three ancillas.
We can see that 
the success rates for $|+\rangle$ and $|-\rangle$ are asymmetric. 
It turns out that the maximum success rate is obtained for $\alpha = 1/3$, and is about $P_{s+ 4, \alpha} = 0.358$ for measuring $|+\rangle$,
but this gives a $0.0037$ success rate for measuring $|-\rangle$, so the average success rate, assuming equal priors for both $|\pm\rangle$, is therefore only about $18.10\%$..
For $\alpha=\pm 1$, we obtain a success rate of only
$0.0544$ for measuring $|-\rangle$, and 0 for measuring $|+\rangle$.
In short, the performance of the 4-component Green Machine with three coherent state ancillas is much lower than a balanced beam splitter with just one coherent state ancilla.

Looking at the falling trend for the success rate as we go from the $n=2$ case to the $n=4$ one, we do not explicitly consider larger Green Machines with coherent state ancillas beyond the $n=4$ case, but it is a straightforward exercise to generalize the success rate formula (\ref{4-component-GM-coherent-state-success-rate}) and obtain
\be
P_{s\pm, n, \alpha}
= 2\exp(-(n-1)|\alpha|^2)
\sum_{i_1, i_2,,\ldots i_n=0, i_1/(n-1)-i_2-i_3-\ldots -i_n =\pm \alpha}^\infty
\frac{(n-1)^{2i_1}}{i_1! i_2! \ldots i_n!}
(|\alpha|^2/n)^{i_1 +i_2+\ldots + i_n}.
\ee 
Here, we are using $i_1, i_2,\ldots i_n$ to label the numbers of photons in modes $1, 2,\ldots , n$, respectively, so the subscripts are part of the different symbols denoting the photon counts in the different modes, and not the mode labels. We are switching to this notation instead of using separate letters as we did for the 2 and 4 component cases to account for the general number of modes.

\end{document}